\documentclass[a4paper,11pt]{article}
\usepackage{pos}
\usepackage{defs}
\author*[a]{Mattia~Dalla~Brida}
\author[b]{Roman H{\"o}llwieser} 
\author[b]{Francesco Knechtli}
\author[b]{Tomasz Korzec}
\author[d]{Alberto~Ramos} 
\author*[e]{Stefan~Sint}
\author[c,f]{Rainer~Sommer}

\affiliation[a]{Theoretical Physics Department, CERN, 1211 Geneva 23, Switzerland}
\affiliation[b]{Department of Physics, University of Wuppertal, Gau{\ss}strasse 20, 42119 Germany}
\affiliation[c]{John von Neumann Institute for Computing (NIC), DESY, Platanenallee~6, 15738~Zeuthen, Germany}
\affiliation[d]{Instituto de F\'isica Corpuscular (IFIC), CSIC-Universitat de Valencia, 46071, Valencia, Spain}
\affiliation[e]{School of Mathematics and Hamilton Mathematics Institute, Trinity College Dublin, Dublin 2, Ireland}
\affiliation[f]{Institut~f\"ur~Physik, Humboldt-Universit\"at~zu~Berlin, Newtonstr.~15, 12489~Berlin, Germany}

\emailAdd{mattia.dalla.brida@cern.ch}
\emailAdd{sint@maths.tcd.ie}

\title{A non-perturbative determination of $\bg$}

\abstract{Close to the continuum limit, lattice QCD with mass-degenerate Wilson quarks can be described by Symanzik's effective continuum action, 
which contains the dimension 5 operator, $m \tr(F_{\mu\nu}F_{\mu\nu})$. 
Its effect can be eliminated by an O($a\mq$) rescaling of the bare lattice coupling constant.
Until recently, the corresponding improvement coefficient, $\bg$, was only known perturbatively to 1-loop order and 
an estimate of the remaining uncertainty is the dominant systematic error in the ALPHA collaboration's 
recent determination of $\alpha_s(m_Z)$ with the decoupling method~\cite{DallaBrida:2022eua}. 
To remove this error we have determined $\bg$ non-perturbatively for the corresponding 
parameter range. We here briefly review improvement conditions for $\bg$, perform a perturbative test and
report on our non-perturbative results for $\bg$.\\

{\vspace*{3mm}\hfill CERN--TH--2024--004}}

\FullConference{The 40th International Symposium on Lattice Field Theory (Lattice 2023)\\
July 31st - August 4th, 2023\\
Fermi National Accelerator Laboratory\\}

\begin{document}
\maketitle

\section{On-shell O($a$) improvement and the r\^ole of $\bg$}

The leading cutoff effects in lattice QCD with Wilson quarks are linear in the lattice spacing $a$, 
due to the explicit breaking of all chiral symmetries by the Wilson term. 
The required counterterms can be classified in Symanzik's effective continuum 
theory by their mass dimensions and lattice symmetries. Including lattice discretizations of these counterterms 
in the action with appropriately tuned improvement coefficients then allows us, at least in principle, 
to completely eliminate O($a$) effects from on-shell quantities such as hadronic masses and 
energies~(cf.~\cite{Luscher:1996sc} and references therein).
Given that O($a$) effects arise from chiral symmetry breaking, one
would expect that chiral Ward identities can be used as improvement conditions 
to determine the counterterm coefficients. For instance, with massless quarks, 
the use of the simplest chiral Ward identity, the PCAC relation, leads to the determination of the coefficient, $\csw$, 
of the Sheikholeslami-Wohlert term and the coefficient $\ca$,
required to improve the axial current. As a result, the uncertainty 
of the massless limit reduces to O($a^2$)~\cite{Luscher:1996sc,Luscher:1996ug}.
With $\Nf$ mass degenerate Wilson quarks, the only additional O($a$) counterterms to the action
take the continuum form $m \tr(F_{\mu\nu}F_{\mu\nu})$ and $m^2 \psibar\psi$, 
and can be implemented on the lattice by re-scaling the bare lattice coupling, $g_0^2$, 
and the bare subtracted quark mass, $\mq = m_0-\mcr$,
\begin{equation}
     \g02t= g_0^2(1+\bg(g_0^2) a\mq), \qquad \mqt = \mq(1+\bm(g_0^2)a\mq)\,.
\end{equation}
We here focus on $\bg$, which is known perturbatively to 1-loop order~\cite{Sint:1995ch},
\begin{equation}
   \bg = \bg^{(1)} g_0^2 +\rmO(g_0^4), \qquad \bg^{(1)} = 0.01200\times\Nf\,,
   \label{eq:bgpert}
\end{equation}
and plays a central r\^ole for consistent O($a$) improvement. It is needed to keep $\g02t$ constant, 
if the quark mass is to be varied at constant lattice spacing. Furthermore, 
renormalization factors in mass-independent renormalization schemes depend on $\g02t$ rather than $g_0^2$.
Perturbative estimates of $\bg$ are fine for small $a\mq$ values, but this
is not the case e.g.~for the charm quark in typical simulations in the hadronic regime.
Our own motivation to compute $\bg$ non-perturbatively originates from the recent determination of
the $\Lambda$-parameter for 3-flavour QCD. There, a fictitious mass-degenerate
triplet of quarks is used to establish a precise connection between renormalized couplings
in 3-flavour QCD and pure gauge theory, respectively~\cite{DallaBrida:2019mqg,DallaBrida:2022eua}. 
The latter is the effective theory in the decoupling limit of infinite quark mass.
In practice, one traces a renormalized QCD coupling for a wide range of quark masses, while
keeping cutoff effects under control. Pure gauge theory calculations can thus be leveraged to obtain
one of the most accurate determinations of the QCD $\Lambda$-parameter and of $\alpha_s(m_Z)$ to date~\cite{DallaBrida:2022eua},
\begin{equation}
 \Lambda^{(3)}_\msbar = 336(10)(6)_{\bg}(3)_{{\hat\Gamma_m}} \MeV = 336(12) \MeV  
 \quad \Rightarrow \quad \alpha_s(m_Z) = 0.11823(84)\,.
  \label{eq:Lambda}
\end{equation}
The quoted systematic error due to $\bg$ was estimated by assuming 
a 100 percent uncertainty on the perturbative result, Eq.~(\ref{eq:bgpert}), 
and this is currently the dominant systematic error. 
We here report on our recent non-perturbatve results for $\bg$~\cite{bg-paper}, aimed at removing 
this error in Eq.~(\ref{eq:Lambda}). Controlling the decoupling limit of large mass $\mq$ while keeping $a\mq$ reasonably
small (values up to $a\mq =0.3-0.4$ were used in the analysis), means that the lattice spacings 
are significantly smaller than in typical hadronic simulations e.g.~by~CLS~\cite{Bruno:2014jqa,cls:status}.

\section{Improvement condition for $\bg$}

We distinguish two strategies to determine $\bg$. The first one is based on chiral Ward identities
and the observation that the O($a$) improvement of the quark bilinear, 
flavour singlet scalar density can be related to $\bg$~\cite{Bhattacharya:2005rb}. 
For a detailed derivation, which also corrects 
an oversight in~\cite{Bhattacharya:2005rb}, we refer to our paper~\cite{bg-paper}.
The second strategy uses restoration of chiral symmetry in a physically small volume of linear dimension $L$,
which implies analyticity of observables in the quark mass~\cite{bg-paper}. Given a gluonic observable $O_g$,
one may thus expand in powers of $z=M L$,
\begin{equation}
   \langle O_g\rangle = 
   \langle O_g\rangle_{z=0} 
   + z \left.\dfrac{\partial\langle O_g \rangle}{\partial z}\right\vert_{z=0}   + \rmO(z^2) 
\end{equation}
where $M$ denotes the RGI quark mass. With a hyper-torus topology and some variant of periodic boundary conditions for all fields,
it can be shown that the physical quark mass dependence must be of O($z^2$). Hence, the term linear in $z$ must be due
to lattice artefacts arising from an incorrect tuning of $\bg$, and 
an improvement condition for $\bg$ is obtained by imposing
\begin{equation}
  \left.\dfrac{\partial\langle O_g \rangle}{\partial z}\right\vert_{z=0} =0\,.
  \label{eq:bg-impcond}
\end{equation}
Assuming $O_g$ to be dimensionless, one may change variables from $(z,\Lambda L,  L/a)$ to $(a\mq,g_0^2, L/a)$ 
and obtain~\cite{bg-paper},
\begin{equation}
  \bg(g_0^2)=\left.\dfrac{\partial \langle O_g \rangle}{\partial a\mq}\right\vert_{g_0^2,\mq=0} 
    \times \left[ \left. g_0^2 \dfrac{\partial \langle O_g \rangle}{\partial g_0^2} \right\vert_{\mq=0} \right]^{-1}\,.
  \label{eq:bg}
\end{equation}

\section{Perturbation Theory}
It is instructive to test this improvement condition in perturbation theory. 
We consider $\langle O_g\rangle = 1/\bar{g}^2$, the SF coupling~\cite{Luscher:1992an,Sint:1995ch},
but with $\chi$SF boundary conditions, which, unlike standard SF boundary conditions, 
do not generate an O($z$) term at the boundaries~\cite{Sint:2010eh,DallaBrida:2016smt}. 
We set the boundary counterterm coefficients $z_{\rm f}$, $\ct$ and $d_{\rm s}$ to their correct 
tree-level values~\cite{DallaBrida:2016smt} and 
use the tree-level O($a$) improved action with $\csw=1$.
We expand perturbatively,
\begin{equation}
   \bar{g}^2(L,z)  = g_0^2 + p_1(L/a,z) g_0^4 + \rmO(g_0^6) = \g02t + \left(p_1(L/a,z)- a\mq\bg^{(1)}\right) \tilde{g}_0^4 + \rmO(\tilde{g}_0^6)\,,
\end{equation}
where we define $z=\mr L$ with some renormalized quark
mass, $\mr$. To lowest perturbative order we have $\mr= \mqt=\mq(1 +\rmO(\mq))$. 
The asymptotic expansion for small $a/L$ takes the form,
\begin{equation}
  p_1 \;\sim \; r_0(z) + 2b_0(z)\ln(L/a) + \dfrac{a}{L} r_1(z) + \rmO(a^2)\,,
\end{equation}
where $b_0$ is the one-loop coefficient of the $\beta$-function, normalized such that $b_0(0) = (11-\frac23 \Nf)/16\pi^2$.
With $\chi$SF b.c.'s, $r_0$ and $b_0$ are even functions of the quark mass, so that their derivatives at $z=0$ must vanish. 
We have anticipated the absence, due to bulk O($a$) improvement, 
of a term $\propto (a/L)\times \ln(L/a)$. Complete tree-level boundary O($a$) improvement implies that $r_1(0)=0$.
The only remaining O($a$) effect is given by the quark mass dependence of $r_1$ and
is cancelled by $\bg$.\footnote{The boundary O($a\mq$) counterterm obtained by the replacement 
$z_{\rm f} \rightarrow z_{\rm f}(1+b_{\rm f}a\mq)$, contributes at O($a^2$) to $p_1$ and at O($a$) 
to the $\bg$-estimate.}

The improvement condition, Eq.~(\ref{eq:bg-impcond}), requires to differentiate
with respect to $z$ at $z=0$, which leads to,
\begin{equation}
\left.\dfrac{\partial \bar{g}^2}{\partial z}\right\vert_{z=0} =  \dfrac{a}{L} \left(r_1'(0)-\bg^{(1)}\right) \tilde{g}_0^4 + \rmO(a^2,\tilde{g}_0^6)  = 0 
 \quad \Rightarrow \quad \bg^{(1)} = r_1'(0)\,.
\end{equation}
We have checked that this reproduces the known one-loop result, Eq.~(\ref{eq:bgpert}).
We can also test Eq.~(\ref{eq:bg}) for a given lattice size. We have\footnote{We here differentiate the inverse observable,
$\gbar^2=1/\langle O_g\rangle$; the additional factor cancels in the ratio for $\bg$.}
\begin{equation}
    \left.\dfrac{\partial \bar{g}^2}{\partial a\mq}\right\vert_{g_0^2,\mq=0} =  g_0^4\left.\dfrac{\partial  p_1}{\partial a\mq}\right\vert_{\mq=0} + \rmO(g_0^6)\,,\qquad  
    \left.g_0^2\dfrac{\partial \bar{g}^2}{\partial g_0^2}\right\vert_{\mq=0} = g_0^2 + 2p_1g_0^4 + \rmO(g_0^6)\,.
\end{equation}
We thus obtain estimates for the 1-loop coefficient at fixed $L/a$,
\begin{equation}
 \bg^{(1),\text{est}} = \left.\dfrac{\partial  p_1}{\partial a\mq}\right\vert_{\mq=0}
= r_1'(0) + \rmO(a/L)\,.
\label{eq:bg-est}
\end{equation}
Using the standard SF coupling with $\chi$SF b.c.'s, $m_0=0$, $\theta=\pi/5$, a few numerical results are
\begin{equation}
    \left.\dfrac{\bg^{(1),\text{est}}}{0.0120\Nf}\right\vert_{L/a}  = 0.899,\,0.913,\,0.924,\,0.940,\, 0.950, 
    \quad\text{for}\quad L/a = 16,24,32,48,64\,.   
\end{equation}
The leading lattice effects are indeed $\propto a/L$, including an $(a/L)\times \ln(L/a)$ term.
In Eq.~(\ref{eq:bg-est}) we have again used,
\begin{equation}
\left.\dfrac{L}{a}\left(r_0'(z) + 2b'_0(z)\ln(L/a)\right)\right\vert_{z=0} = 0\,, 
\quad \text{due to} \quad r_0'(z), b'_0(z) = \rmO(z)\,.
\end{equation}
At finite $L/a$, the chiral limit is not unambiguously defined. Setting $z=0$
means $\mq=0$ up to an O($a^2$) uncertainty. This implies that the O($a$) improved axial current (with counterterm
$\propto \ca$) must be used to obtain an O($a$) improved critical mass from the PCAC relation. 
At 1-loop order, we have tested this effect by setting $m_0 \propto a/L^2$, i.e. by adding
an artificial O($a$) term to the critical mass, which then produces an O($1$) shift in $\bg^{(1),\text{est}}$, as expected.

Unfortunately, $\chi$SF b.c.'s are only available for even $\Nf$. For odd flavour numbers, a mixed set-up with a 
single massless ``spectator quark'' satisfying SF boundary conditions~\cite{DallaBrida:2018tpn}, while a theoretical possibility,
looks less attractive, and some hyper-torus topology with periodic or twisted periodic (for $\Nf=3$) 
boundary conditions is clearly preferable.

\section{Non-perturbative results}

To determine $\bg$ non-perturbatively for $\Nf=3$ O($a$) improved Wilson quarks~\cite{Bulava:2013cta} 
and the L\"uscher-Weisz gauge action~\cite{Luscher:1984xn}, 
we chose a line of constant physics (LCP) defined by $z=0$ and the physical scale $L$ defined 
implicitly by $\bar{g}^2_{\rm GF}(L) =3.949$~(see ref.~\cite{DallaBrida:2016kgh} for the coupling definition), 
corresponding to $L\approx 0.25\,\text{fm}$~\cite{DallaBrida:2022eua}.
For a given lattice size, $L/a$, the LCP implies values for $\beta=6/g_0^2$ and the critical mass, 
$a\mcr(g_0^2)$. With $L/a = 12,\ldots,48$ this leads to $\beta$-values in the range $\beta\in [4.3,5.17]$.
Periodic and anti-periodic boundary conditions in all directions were imposed for 
the gauge and quark fields, respectively. This allowed us to simulate around the chiral limit without the need
for chiral extrapolations. We looked at two sets of renormalized and O($a$) improved gluonic observables, 
first the gluon action density at finite flow-time, $t$,
\begin{equation}
   \sigma(c) = \sum_{\mu,\nu = 0}^3
    \dfrac{t^2\,\langle {\rm tr}\left\{G_{\mu\nu}(t,x)G_{\mu\nu}(t,x)\right\}\delta_{Q(t),0} 
    \rangle}{\langle \delta_{Q(t),0}\rangle}\Bigg|_{8t=c^2L^2}\,,
 \label{eq:GF}
\end{equation}
where the projection onto the zero topological charge sector is part of the definition of $\sigma(c)$.
Secondly, we measured Creutz ratios, defined from expectation values of rectangular Wilson loops, $W(R,T)$, by
\begin{equation}
    \chi(R,T)=-\tilde\partial_R \tilde\partial_T \log(W(R,T))\,,
    \label{eq:chi1}
\end{equation}
with symmetric lattice $R,T$-derivatives. Again, a projection to the $Q=0$ sector 
is implicit in the definition of $W(R,T)$. To complete the definition of the LCP we still need to
fix the flow time and the dimensions of the Wilson loops in units of $L$. After some
exploration in parameter space, our main (dimensionless) observables 
were chosen as $\sigma(c=0.18)$ and $\hat\chi = R^2\chi(R,R)|_{R=L/4}$. 

Estimates of $\bg$ by Eq.~(\ref{eq:bg}) require the derivatives of $\sigma(c), \hat\chi$, 
with respect to the bare parameters. These were implemented by tracing their $\mq$-dependence 
at fixed $\beta$ and their $g_0^2$-dependence at $\mq=0$, respectively. Various checks were performed, 
in particular we made sure that higher order terms in $\mq$ did not impact on our extraction 
of the linear $\mq$ dependence required for $\bg$. Since $\bg$ is obtained from a 
cutoff effect that vanishes proportionally to $a/L$, it gets increasingly
difficult to obtain direct measurements on the larger lattices. We decided to strictly follow the LCP
for lattice sizes from $L/a=12$ to $L/a=24$. At the $L/a=32$ point of the LCP we then performed some 
checks on the $L/a$-dependence of the $\bg$-estimates. We concluded that the difference was 
too small to be resolved, allowing us to continue on with $L/a=24$ for this and the higher $\beta$-values.
We also checked the sensitivity of our $\bg$-estimates to the precise definition of
the LCP, by studying their dependence on $t$ and $R$. It turns out that these estimates are very stable
once $\sqrt{8t}/a, R/a \ge 3$ which is satisfied for all $\bg$-estimates from $\hat\chi$; 
with $\sigma(0.18)$ this holds for all lattice sizes except $L/a=12,16$ where
the $t$-dependence is visible but still rather mild. 

\begin{figure}[htbp]
\centering
\includegraphics[width=.49\linewidth]{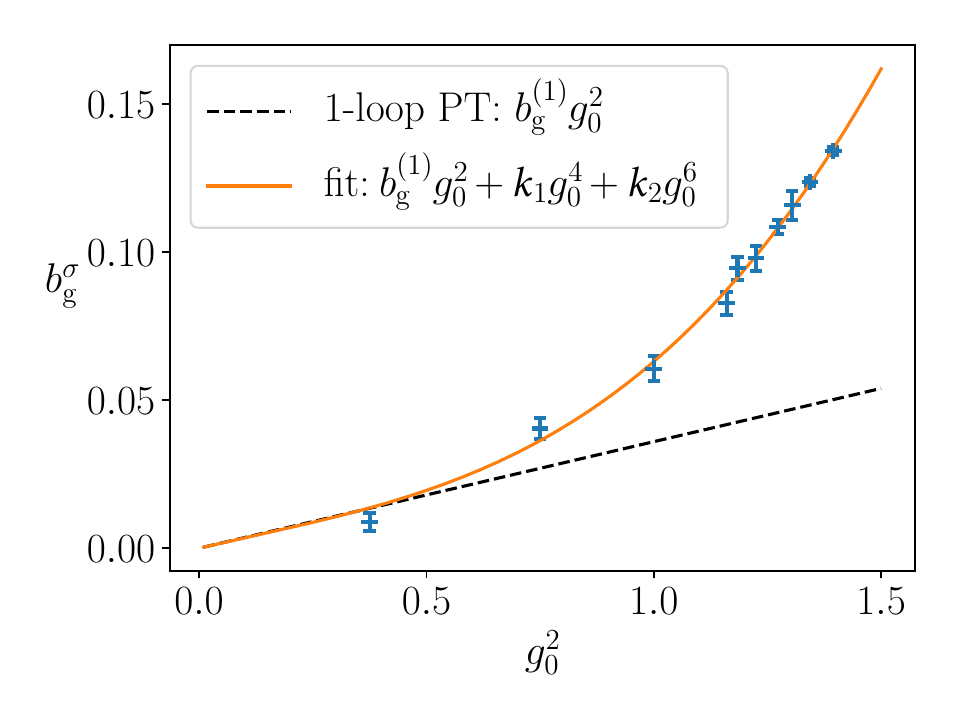}
\includegraphics[width=.49\linewidth]{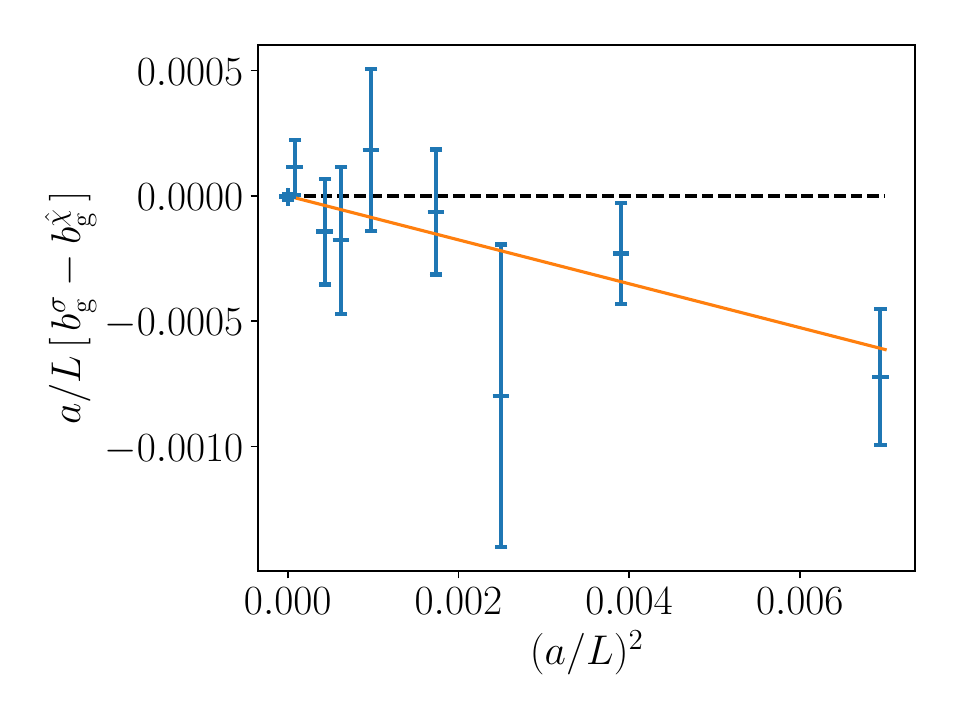}
\caption{Final result for $\bg$ from $\sigma(0.18)$ (left) and ambiguity $[\bg^\sigma-\bg^{\hat\chi}]\,\times \frac aL$ (right). 
The extra factor $a/L$ is added because $\bg$ always enters into observables with  an explicit factor of $a$; 
we also include a linear fit to all points constrained to go to zero.}
        \label{fig:dbg}
\end{figure}

We also produced data at $\beta=6,8,16$, in order to compare with perturbation theory.
The results for $\bg$ obtained from $\sigma(c=0.18)$ are collected in Figure~\ref{fig:dbg} (left panel) 
and smoothly connect the perturbative and non-perturbative regimes, even though
the deviation from perturbation theory becomes quite large at the lowest $\beta$-values. 
In order to check for consistency we looked at the difference to the $\bg$-estimates from $\hat\chi$.
We found that this O($a$)-ambiguity is very small (see right panel of Figure~\ref{fig:dbg}), 
and compatible with O($a$). 
We have performed different fits to the data for $\bg$ from $\sigma(0.18)$. 
Our preferred 2-parameter representation for $\beta\ge 4.3$ is given by
\begin{equation}
 \bg^\sigma = 0.036\,g_0^2 -0.0151\,g_0^4 + 0.0424\, g_0^6\,,
\label{eq:res}
\end{equation}
where the first coefficient is constrained to the perturbative result for $\Nf=3$, Eq.~(\ref{eq:bgpert}).
The overall uncertainty from comparing to different fits is $1.1\times 10^{-3}$ and 
ca.~$4\times 10^{-4}$ in the range $\beta \in [4.3,5.17]$.

\section{Conclusions}

We have reported on our recent non-perturbative results for the improvement coefficient $\bg$,
as needed for the study of the decoupling limit in~\cite{DallaBrida:2019mqg,DallaBrida:2022eua}.
The improvement condition relies on the restoration of chiral symmetry in a physically small volume
and passes a perturbative test to first non-trivial order.
Following a line of constant physics, we have looked at the action density at finite gradient flow time 
and Creutz ratios, in order to test the dependence of $\bg$-estimates on the choice of observable. 
We find remarkable stability of our results, which are well-represented by a 2-parameter fit, Eq.~(\ref{eq:res}).
The re-analysis of the data from the decoupling project for the QCD $\Lambda$-parameter and $\alpha_s(m_Z)$ 
is currently underway.

\section*{Acknowledgements}
The work is supported by the German Research Foundation (DFG) research unit 
FOR5269 ``Future methods for studying confined gluons in QCD''.  
RH was supported by the programme ``Netzwerke 2021'', an initiative of the Ministry 
of Culture and Science of the State of Northrhine Westphalia, in the NRW-FAIR network, 
funding code NW21-024-A. AR acknowledges ﬁnancial support from
the Generalitat Valenciana (genT program CIDEGENT/2019/040) and the Spanish Ministerio de
Ciencia e Innovacion (PID2020-113644GB-I00). SS and RS acknowledge funding by the H2020 program in the  
{\em Europlex} training network, grant agreement No. 813942. 
Generous computing resources were supplied by the North-German Supercomputing Alliance 
(HLRN, project bep00072) and by the John von Neumann Institute for Computing (NIC) at DESY, 
Zeuthen. The authors are grateful for the hospitality extended to them at CERN during the
initial stage of this project. 
We thank our colleagues in the ALPHA-collaboration for valuable discussions.

\bibliographystyle{utphys}

\providecommand{\href}[2]{#2}\begingroup\raggedright\endgroup

\end{document}